\documentclass[aps,pl,amsmath,amssymb,twocolumn,letterpaper,%
         footinbib,bibnotes,showpacs,reprint,balancelastpage,raggedbottom,
         citeautoscript,floatfix,superscriptaddress]{revtex4-1}
\usepackage{graphicx}
\usepackage{bm}
\usepackage{dcolumn}
 \usepackage{xfrac}
\usepackage{xcolor}
\usepackage{microtype}
\usepackage{hyperref}

\begin{document}
\title{The influence of Coulomb Correlations and Spin-Orbit Coupling in the electronic structure of double perovskites Sr$_2$XOsO$_6$ (X$=$Sc, Mg) }

\author{G.\ Giovannetti}
\affiliation{Institute for Theoretical Solid State Physics, IFW-Dresden, PF 270116, 01171 Dresden, Germany}

\begin{abstract}
We investigate the antiferromagnetic insulating state of the recently discovered double perovskites Sr$_2$XOsO$_6$ (X$=$Sc, Mg) by using ab-initio calculations (based on Density Functional Theory and Dynamical Mean-Field Theory) to elucidate the interplay between electronic correlations and spin-orbit coupling. 
The structural details of Sr$_2$XOsO$_6$ (X$=$Sc, Mg) induce band narrowing effects which enhance local electronic correlations.
The half-filled $5d^3$ orbitals of Os in Sr$_2$ScOsO$_6$ fall into a magnetically ordered correlated regime, which is slightly affected and reduced by the spin-orbit coupling.
The electronic configuration $5d^2$ of Os in Sr$_2$MgOsO$_6$ responses differently to electronic correlations promoting a less localized state than Sr$_2$ScOsO$_6$ at the same strength of electronic interactions. We find that the inclusion of spin-orbit coupling drives Sr$_2$MgOsO$_6$ toward insulating behaviour and promotes a large tendency in formation of orbital magnetization antiparallel to the spin moment. The formation of the AFM state is linked to the evidence of correlated Hubbard bands in the paramagnetic solution of Sr$_2$XOsO$_6$ (X$=$Sc, Mg).
\end{abstract}

\maketitle

\section{Introduction} 

The double perovskites are a fascinating playground of different properties such as half-metallicity \cite{Retuerto,Philipp}, high temperature ferrimagnetism \cite{Feng,Krockenberger}, ferroelectricity \cite{Fukushima,Kumar,Retuerto1} and a rich variety of magnetic interactions \cite{Ou,Morrow1}.
The choice of A, B and B$^{'}$ in their chemical formula A$_2$BB$^{'}$O$_6$, having transition metal ions B and B$^{'}$ residing on the two sublattices of a three-dimensional cubic lattice, avoids related difficulties in designing different suitable electronic properties for several potential applications.

Recently double perovskites Sr$_2$XOsO$_6$ (X$=$Sc, Mg) hosting a single magnetic ions X have been considered for their high magnetic transition temperatures.

The magnetization measurements of Sr$_2$ScOsO$_6$ indicate an antiferromagnetic transition to type-I magnetic ordering (the two Os sites in the unit cell are coupled antiferromagnetically with spins oriented in the ab-plane) at T$_N$ = 92 K, one of the highest transition temperatures of any double perovskite \cite{Sr2ScOsO6}. Using both neutron and x-ray powder diffraction the crystal structure of Sr$_2$ScOsO$_6$ has been refined as monoclinic with symmetry $P2_1/n$ at all the temperatures \cite{Sr2ScOsO6}. 

The compound Sr$_2$MgOsO$_6$ represents another rare example of double perovskites with T$_N$ = 100 K, higher than Sr$_2$ScOsO$_6$, and with the same type-I magnetic ordering \cite{Sr2MgOsO6}.
Its crystal symmetry has been refined in the tetragonal symmetry $I4/m$ and the temperature dependence of the resistivity of the polycrystalline Sr$_2$MgOsO$_6$ indicates semiconductor-like behavior \cite{Sr2MgOsO6}.

In both the compounds the inverse susceptibility versus temperature shows clearly Curie-Weiss behaviour signaling the existence of local magnetic moments \cite{Sr2ScOsO6,Sr2MgOsO6}. 

The differences in the crystal structure of Sr$_2$ScOsO$_6$ and Sr$_2$MgOsO$_6$ are rather small and related to the crystal group symmetries in which the tilting between neighboring layers of octahedral environment are $a^{-}a^{-}c^{+}$ and $a^{0}a^{0}c^{−}$ (in the Glazer’s notations).

In terms of electronic structure the double perovskite Sr$_2$ScOsO$_6$ and Sr$_2$MgOsO$_6$ are expected to have the transition metal Os ions in nominal 5$d^{3}$ and 5$d^{2}$ electronic configurations, which are combined with the ones of Sc$^{3+}$ and Mg$^{2+}$ ions. For such electronic configurations the effect of spin-orbit coupling (SOC) in 5d orbitals is expected to be rather different \cite{ChenBalents,Pardo,Lee2,Gangopadhyay} as well as the role of electronic correlations \cite{GeorgesAnnuRev,DeMedici}.

The explanation of the presence of localized electrons in Sr$_2$XOsO$_6$ (X$=$Sc, Mg), which can host high magnetic ordering temperatures \cite{Sr2ScOsO6,Sr2MgOsO6}, it is hidden in the details of the interplay of SOC and electronic correlations in the Os 5d electrons \cite{Sr2ScOsO6,Sr2MgOsO6}.

In this work using a set of ab-initio calculations we provide evidences of a correlated Mott magnetic state in recently discovered double perovskites Sr$_2$XOsO$_6$ (X$=$Sc, Mg). We investigate the electronic structure of Sr$_2$XOsO$_6$ (X$=$Sc, Mg) pointing out how the interplay of SOC and electronic correlations favors the inset of localization in the 5d manifold and as a consequence the magnetic ordering.

\section{Calculation Details}

We perform first-principles density functional calculations within the Local Density Approximation approximation (LDA) \cite{LDA} as
implemented in the Vienna {\it Ab initio} Simulation Package (VASP) \cite{VASP} with the projector augmented wave (PAW) method \cite{PAW} to treat the core and valence electrons.
A kinetic cutoff energy of 500\,eV is used to expand the wavefunctions and a $\Gamma$ centered  6$\times$6$\times$4 $k$-point mesh combined with the tetrahedron and Gaussian methods is used for Brillouin zone integrations. 

\begin{figure}[]
\vspace{0.1cm}
\includegraphics[width=0.875\columnwidth,angle=-0]{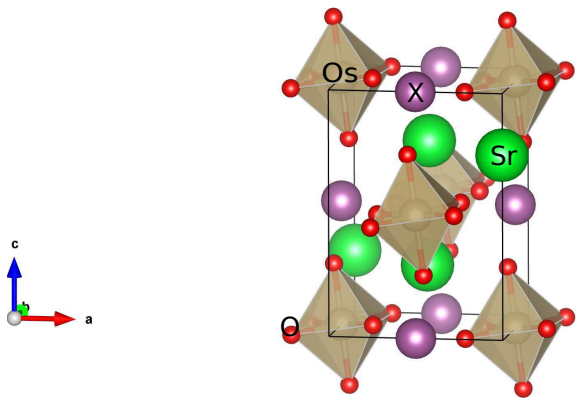} \\
\vspace{0.1cm}
\caption{(Color online) Schematic view of the unit cell of Sr$_2$XOsO$_6$ (X$=$Sc, Mg). The Os octahedra are highlighted by a polyhedral shape.}
\label{fig0}
\end{figure}

The ionic and cell parameters of Sr$_2$XOsO$_6$ (X$=$Sc, Mg) are fixed to the experimental ones \cite{Sr2ScOsO6,Sr2MgOsO6} (see Fig. \ref{fig0} for a schematic view of the unit cell).

Since it is well known that the LDA often underestimates the effects of electronic correlations in systems with $d$ orbitals we also calculate electronic properties within the rotationally invariant LDA+$U$ method introduced by Liechtenstein et al. \cite{DFTplusU} which requires two parameters, the Hubbard parameter $U$ and the exchange interaction J$_h$. 
Performing magnetic spin polarized calculations in the AFM type-I order within this scheme we vary the magnitude of the Hubbard parameter $U$ between 0 and 4\,eV for the Os $d$-states. Note that the standard spin-polarized LDA corresponds to $U=$J$_h$=0\,eV. We explicity include the SOC in the LDA and LDA+U calculations.

To treat local electronic correlations beyond mean-field theory and to better understand the effects of electronic correlations in double perovskites Sr$_2$XOsO$_6$ (X$=$Sc, Mg) we combine DFT and Dynamical Mean-Field Theory (DMFT)\cite{DMFT} and we focus mainly on the paramagnetic state, where magnetic ordering in inhibited. From our point of view, the choice of paramagnetic solutions allows us to characterize the parent state from which AFM establishes due to the presence of localized electrons. 

In our implementation of the LDA+DMFT and LDA+SOC+DMFT approach, we first construct maximally localized Wannier orbitals (using Wannier90 code \cite{wannier90}) for the $5d$ Os orbitals over the energy range spanned by the Os $t_{2g}$ from the LDA band structure without and with including SOC. 
When the SOC is included we use the numerical J,J$_z$ basis, which diagonalizes the onsite part of the Hamiltonian, to perform the calculations (in the following we refer to U as the matrix describing such rotation of the basis set) otherwise the Os $t_{2g}$ manifold is used.
Then we solve the self-consistent impurity model using Exact Diagonalization\cite{Caffarel,Capone}, in which the impurity model is described with a number of levels $N_s=12$ \cite{Liebsch,2322,BaCrO3GG}, and diagonalized by a parallel Arnoldi algorithm \cite{ARPACK}.

In the LDA+SOC+DMFT the Coulomb interaction was treated as a parametrized form in which the interactions in the J,J$_z$ basis are calculated from the ones used in the LDA+DMFT between the $t_{2g}$ orbitals in the density-density approximation as done in Ref. \cite{AritaSr2IrO4}.
This corresponds to use the same matrix U defined above to rotate the interaction matrix in density-density approximation as defined in the $t_{2g}$ orbitals (L,L$_z$ basis set).
For LDA+DMFT calculations we consider as terms of interactions U (between electrons occupying same orbital but opposite spin channel), U$^{'}$=U-2J$_h$ (between electrons occupying different orbitals but same spin channel), U$^{''}$=U-3J$_h$ (between electrons occupying different orbitals but opposite spin channel).
The precise value of U and J$_h$ in 5d Os orbitals in Sr$_2$XOsO$_6$ (X$=$Sc, Mg) is unknown and then we perform calculation LDA+DMFT and LDA+SOC+DMFT calculations for different values of U at fixed ratio J$_h$/U kept to 0.15 and 0.3.

\section{Results}

\subsection{LDA and LDA+U calculations}
\begin{figure}[]
\vspace{0.1cm}
\includegraphics[width=0.6\columnwidth,angle=-90]{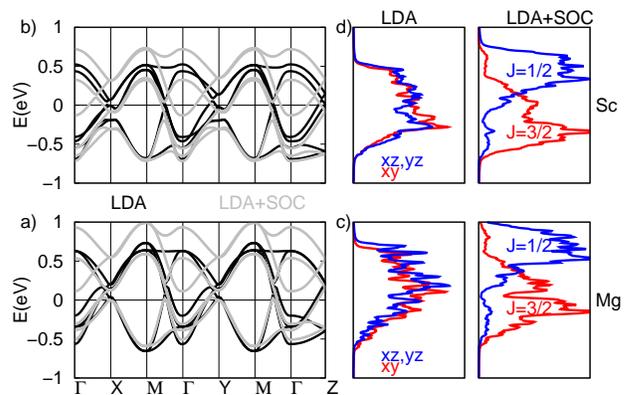} \\
\vspace{0.1cm}
\caption{(Color online) 
Band structure of Sr$_2$MgOsO$_6$ (a)) and Sr$_2$ScOsO$_6$ (b)) within LDA and LDA+SOC scheme along high symmetry direction of the Brillouin zone. The Fermi level is set to zero. The t$_{2g}$ orbital (LDA) and total angular momentum J (LDA+SOC) projected density of states based on Wannier functions for Sr$_2$MgOsO$_6$ (c)) Sr$_2$ScOsO$_6$ (d)).}
\label{fig1}
\end{figure}

We start to determine the band structures of Sr$_2$XOsO$_6$ (X$=$Sc, Mg) by means of DFT paramagnetic calculations as shown in Fig. \ref{fig1}a),b) \cite{notaBZ}.
LDA calculations clearly lead to a metallic solution with a sizable spectral weight at the Fermi level. 
The low-energy contribution to the spectral density is dominated by the bands arising from t$_{2g}$ Os 5d electrons, which are very weakly entangled with oxygen bands lying in energy windows well below the Fermi level.
The octahedral environment splits the Os 5d orbitals into t$_{2g}$ and e$_g$ levels, stabilizing 5$d^{3}$ and 5$d^{2}$ electronic configurations respectively in Sr$_2$ScOsO$_6$ and Sr$_2$MgOsO$_6$. 
The bandwidths W of both double perovskites, related to the Os t$_{2g}$ orbitals, are narrow as consequence of the staggering coordination of Os sites combined with Sc and Mg non-magnetic ions in their crystal structure. The values of W are $\sim$ 1.2 and 1.35 eV respectively for X$=$Sc and Mg. This difference is related to the reduced degree of buckling in the crystal structure of Sr$_2$MgOsO$_6$ compared to the one of Sr$_2$ScOsO$_6$ \cite{Sr2ScOsO6,Sr2MgOsO6}.

The effect of the inclusion SOC in DFT paramagnetic computations is sizeable (see Fig. \ref{fig1}) and it relates to the broadening of the bandwidths of Os t$_{2g}$ orbitals in Sr$_2$XOsO$_6$ (X$=$Sc, Mg).
In our LDA+SOC framework the bandwidths W now increase to $\sim$ 1.45 and 1.6 eV respectively for X$=$Sc and Mg.

\begin{figure}[]
\vspace{0.5cm}
\includegraphics[width=0.6\columnwidth,angle=-90]{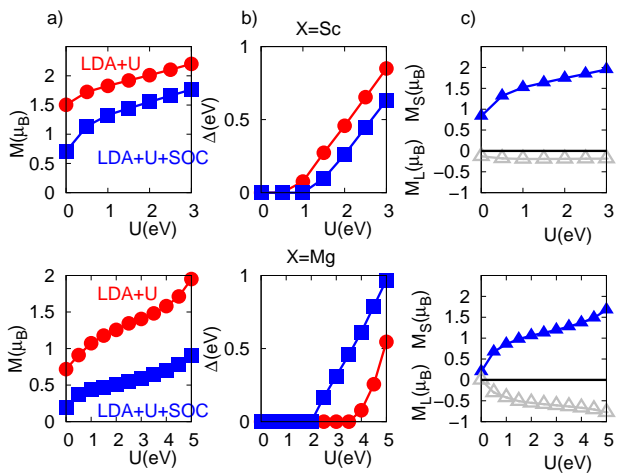} \\
\vspace{0.5cm}
\caption{(Color online)
Magnetization M (a)), charge gap $\Delta$ (b)) as a function of U within LDA+U and LDA+U+SOC calculations for Sr$_2$MgOsO$_6$ and Sr$_2$ScOsO$_6$ in the AFM type-I magnetic structure. 
Orbital M$_L$ and spin M$_S$ magnetization (c))  as a function of U within LDA+U+SOC calculations for Sr$_2$MgOsO$_6$ and Sr$_2$ScOsO$_6$ in the AFM type-I magnetic structure.
}
\label{fig2}
\end{figure}

To investigate the effect of SOC in the electronic structure of the Os t$_{2g}$ orbitals for X$=$Sc and Mg we construct maximally localized Wannier orbitals \cite{wannier90} for the $5d$ Os orbitals over the energy range spanned by the Os $t_{2g}$ states in the LDA and LDA+SOC band structures (see Fig. \ref{fig1}).

In these calculations for Sr$_2$XOsO$_6$ (X$=$Sc, Mg) in the t$_{2g}$ manifold the orbitals d$_{xz}$ and d$_{yz}$ are found degenerate and lower in energy of 0.08 eV respect to the orbital d$_{xy}$ (see Fig. \ref{fig1}c) and d)).
In Fig. \ref{fig1}c) and Fig. \ref{fig1}d) the total angular momentum resolved density of states (evaluated in LDA+SOC approach in the Wannier basis set) clearly shows the formation of a lower J=$\frac{3}{2}$ quartet and an higher J=$\frac{1}{2}$ doublet, which are splitted by $\sim$ 0.5 eV.
The same splitting is also found to be relevant in the electronic structure of the double peroskite Ba$_2$NaOsO$_6$ with d$^1$ configuration \cite{Lee}.
In LDA scheme the orbitals occupancies (calculated from Wannier orbitals) of t$_{2g}$ (xy,xz,yz) orbitals are $\sim$ 0.5 and 0.33 $e$ per spin channel respectively for X$=$Sc and Mg while in LDA+SOC for the set of J=($\frac{3}{2}$,$\frac{1}{2}$) orbitals become $\sim$ (0.7,0.1) and (0.45,0.1) $e$ per spin channel respectively for X$=$Sc and Mg.

To deep our understanding the interplay of the effects related to SOC and local electronic correlations in the AFM state of Sr$_2$XOsO$_6$ (X$=$Sc, Mg) we perform LDA+U and LDA+SOC+U calculations as function of different strengths of electronic correlations in Sr$_2$XOsO$_6$ (X$=$Sc, Mg) in the type-I ordered magnetic structure \cite{Sr2ScOsO6,Sr2MgOsO6,notaAFM}. 

The local magnetization M and charge gap $\Delta$􏰁 as a function of U are reported in Fig. \ref{fig2}a) and \ref{fig2}b) for LDA+U and LDA+SOC+U calculations (J$_h$=0). 
The Os-O hybridization reduces the Os spin moment from the expected 3(2)${\mu}_B$ corresponding to S=3/2(1) for Sr$_2$ScOsO$_6$ (Sr$_2$MgOsO$_6$).
The orbital and spin moments, M$_L$ and spin M$_S$, are antiparallel to each other in both compounds (see Fig. \ref{fig2}c)).
The orbital moment M$_L$ is almost constant in Sr$_2$ScOsO$_6$ while it increases in Sr$_2$MgOsO$_6$ over all the range of U. 
The band gap $\Delta$ correlates strongly with the module of the Os magnetization which increases by increasing U.
As shown in Fig. \ref{fig2} at the optimal value of U=2 and 3 eV we can match the values of Os local magnetization (M) 1.55 and 0.59 ${\mu}_B$ found experimentally respectively for X$=$Sc, Mg \cite{Sr2ScOsO6,Sr2MgOsO6} (with band gaps $\Delta$ of 0.26 and 0.31 eV). The difference in the optimal values of U in Sr$_2$XOsO$_6$ (X$=$Sc, Mg) can be probably related to the different electronic configuration 5d$^3$ (X$=$Sc) and 5d$^2$ (X$=$Mg) but advanced cRPA calculations to evaluate the proper values of U would be really useful to elucidate this point \cite{Sasioglu}.

\begin{figure}[]
\vspace{0.5cm}
\includegraphics[width=0.6\columnwidth,angle=-90]{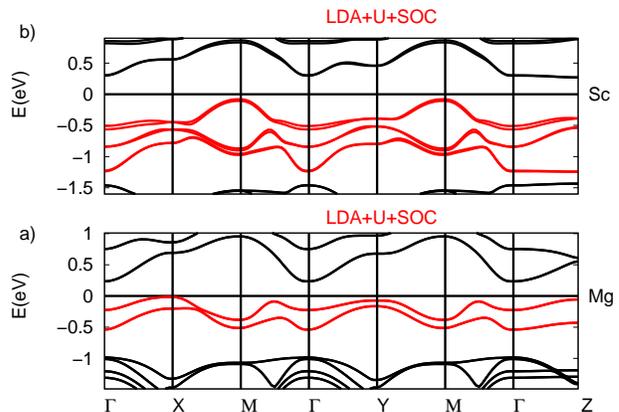} \\
\vspace{0.5cm}
\caption{(Color online) Band structure of Sr$_2$MgOsO$_6$ (a)) (U=3.0 eV) and Sr$_2$ScOsO$_6$ (b)) (U=2.0 eV) within LDA+U+SOC scheme for the AFM type-I magnetic order along high symmetry direction of the Brillouin zone. The Fermi level is set to zero. The two (three) bands of Sr$_2$MgOsO$_6$ (Sr$_2$ScOsO$_6$) that were half filled in the J,J$_z$ basis set are highlighted in red.}
\label{fig5}
\end{figure}

The inclusion of the SOC in Sr$_2$ScOsO$_6$ i) decreases sligthly the Os local magnetization M and ii) increases the critical value of interactions U to reach the insulating AFM state. 
This is a common feature with other half-filled 5d$^3$ Os based materials \cite{LiOsO3GGMC,LiOsO3LiNbO3,Jung}.
In Sr$_2$MgOsO$_6$ the LDA+U+SOC calculations show a large reduction of the i) Os local magnetization and ii) critical value of interactions U to reach the insulating state respect to the LDA+U scheme. 

We mention that different values of Hund's coupling J$_h$ do not change the qualitative picture at level of LDA+U+SOC calculations in the AFM type-I magnetic ordered state.
Indeed for the optimal values of U (U=2 and 3 eV for respectively X$=$Sc, Mg) and J$_h$/U = 0.15 and 0.3 we find 1.51 and 1.47 ${\mu}_B$ for Sr$_2$ScOsO$_6$ and 0.50 and 0.42${\mu}_B$ for Sr$_2$MgOsO$_6$. 

In Fig. \ref{fig5} we show the band structure calculated in the AFM type-I magnetic order for Sr$_2$XOsO$_6$ (X$=$Sc, Mg) within LDA+U+SOC scheme.
For both the compounds we find that two (X=Mg) and three (X=Sc) half-filled bands in the J,J$_z$ basis set split into fully filled lower and unfilled upper Hubbard bands.
Then the combined effects of spin-orbit coupling and on-site Coulomb repulsion result in an AFM Mott insulating state in both compounds.

\subsection{LDA+DMFT and LDA+SOC+DMFT calculations}

\begin{figure}[]
\vspace{0.5cm}
\includegraphics[width=0.6\columnwidth,angle=-90]{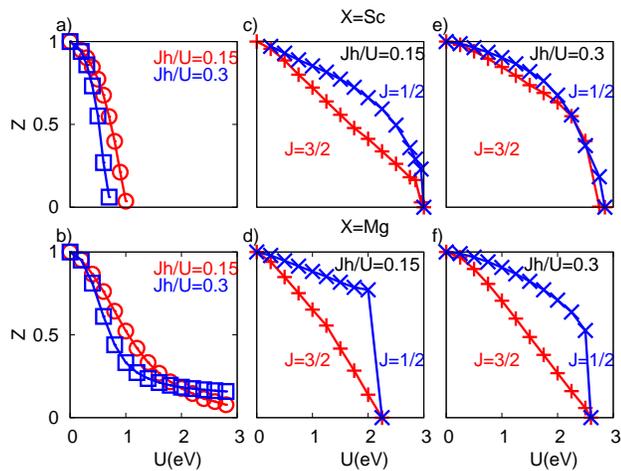} \\
\vspace{0.5cm}
\caption{(Color online) Quasiparticle weight Z of t$_{2g}$ orbitals of Sr$_2$ScOsO$_6$ (a)) and Sr$_2$ScOsO$_6$ (b)) in LDA+DMFT scheme for J$_h$/U=0.15 and 0.3.
Quasiparticle weight Z of J$=\frac{3}{2}$,$\frac{1}{2}$ orbitals for Sr$_2$ScOsO$_6$ (c) for J$_h$/U=0.15 and e) for J$_h$/U=0.3) and Sr$_2$MgOsO$_6$ (d) for J$_h$/U=0.15 and f) for J$_h$/U=0.3) in LDA+SOC+DMFT scheme.}
\label{fig6}
\end{figure}

To further relate to the magnetic scenario found experimentally the presence of electronic correlations and Mott state we perform LDA+DMFT and LDA+SOC+DMFT calculations.
In Fig. \ref{fig6}  we show the quasiparticle weigths in the LDA+DMFT and LDA+SOC+DMFT respectively for Sr$_2$MgOsO$_6$ and Sr$_2$ScOsO$_6$ at different values of the ratio J$_h$/U.

The underlying physics related to the role of electronic correlations of the Os 5d$^2$ (Sr$_2$MgOsO$_6$) and 5d$^3$(Sr$_2$ScOsO$_6$) configurations within our LDA+DMFT scheme (without taking into account SOC) follows the trends already described for model calculations in Ref. \cite{GeorgesAnnuRev,DeMedici} (see \ref{fig6} a) and b)).
In Sr$_2$ScOsO$_6$ the Os t$_{2g}^3$ orbitals are close to degeneracy and half-filled (0.5 $e$ per spin channel), a configuration which is well known to be particularly prone to the effects of electronic correlations due to the Coulomb interactions and Hund's coupling \cite{GeorgesAnnuRev,DeMedici}. The quasiparticle weights Z of all these half-filled Os t$_{2g}$ orbitals decrease by increasing the Hund's coupling in Sr$_2$ScOsO$_6$ (see Fig. \ref{fig6}a)) and we find the critical value of U for the metal-insulator to be $\sim$ 1 eV for both values of the ratio J$_h$/U.

The electronic configuration 5d$^2$ in Sr$_2$MgOsO$_6$, with t$_{2g}$ orbitals close to degeneracy and populated 0.33 $e$ per spin channel, still is affected to a significant extent by electronic correlations \cite{GeorgesAnnuRev,DeMedici}.
We find a two-fold effect of the Hund's coupling: a drastic reduction of the quasiparticle weight Z of all the Os t$_{2g}$ at small strength of electronic interactions and an increase of its critical value to reach the Mott state \cite{GeorgesAnnuRev,DeMedici} (see Fig. \ref{fig6}b)). Although Sr$_2$MgOsO$_6$ is prone to bad metallic behaviour the Mott insulating phase is pushed to large values of the interaction strength U larger than 3 eV.

When SOC is included at level of LDA+SOC+DMFT calculations the physical picture, in terms of localization of quasiparticle weights Z and critical values of U to reach the Mott state, changes dramatically (see Fig.\ref{fig6}c), d), e) and f)) due to the reshape of interband and intraband interactions terms and the loss of the orbital degeneracy in the J,J$_z$ basis set.
The orbital differentiation between J=$\frac{3}{2}$ quartet and J=$\frac{1}{2}$ doublet is larger in Sr$_2$MgOsO$_6$ than in  Sr$_2$ScOsO$_6$. 
The inclusion of the SOC in LDA+SOC+DMFT calculations sets a smaller(larger) value of the strength of Coulomb interactions in Sr$_2$MgOsO$_6$ (Sr$_2$ScOsO$_6$) than the one obtained in LDA+DMFT scheme. 

\begin{figure}[]
\vspace{0.5cm}
\includegraphics[width=0.6\columnwidth,angle=-90]{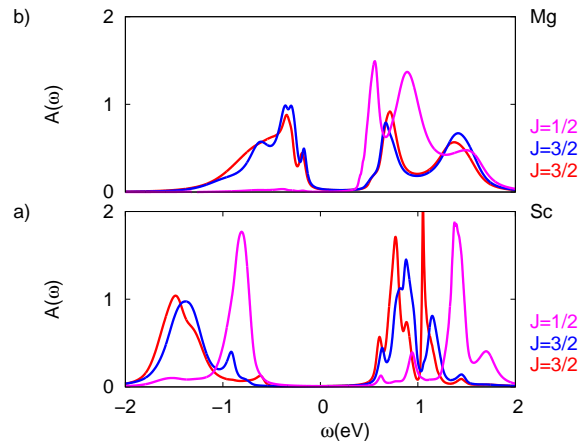} \\
\vspace{0.5cm}
\caption{(Color online) LDA+DMFT orbital J resolved spectral density for Sr$_2$ScOsO$_6$ (a)) for U=3.0 eV and J$_h$/U = 0.3 and Sr$_2$MgOsO$_6$ (b)) for U=2.5 eV and J$_h$/U = 0.15. } 
\label{fig4}
\end{figure}

The effects of the electronic correlations is to drive orbitals occupancies for the set of J=($\frac{3}{2}$,$\frac{1}{2}$) orbitals towards $\sim$ (0.5,0.5) and (0.5,0.0) $e$ per spin channel respectively for X$=$Sc and Mg.
In our computations we find the inset of Mott state is stabilized in Sr$_2$MgOsO$_6$ in correspondence of orbital occupancies of J=$\frac{3}{2}$ close to half-filling and empty J=$\frac{1}{2}$ states while in Sr$_2$ScOsO$_6$ it is obtained when both the J=$\frac{3}{2}$,$\frac{1}{2}$ states are very close half-filled.
The smallest critical values of on-site Coulomb repulsion U are respectively 2.75 eV for X$=$Sc (J$_h$/U=0.3) and 2.25 eV for X$=$Sc (J$_h$/U=0.15).
We observe that the quasiparticle weights Z and the critical values of on-site Coulomb repulsion U decrease(increase) by increasing the Hund's coupling in X$=$Sc(Mg).
This is due to the interplay between Coulomb interactions and the effective crystal-field splitting in the J,J$_z$ basis set which determines the orbital occupancies of J=$\frac{3}{2}$ and J=$\frac{1}{2}$ states and controls consequently their degree of electronic correlations.

In Fig. \ref{fig4} we show the J,J$_z$ basis resolved spectral density in the paramagnetic insulating phase calculated within our LDA+SOC+DMFT scheme \cite{notaPade} for different values of interactions. 
The insulating Mott state is clearly demostrated by the presence of two (X=Mg, J=$\frac{3}{2}$) and three (X=Sc, J=$\frac{3}{2}$,$\frac{1}{2}$) fully filled lower and unfilled upper Hubbard bands in close similarity to what we mentioned in our LDA+SOC+U results.

We remark that the electronic structure of Sr$_2$XOsO$_6$ (X$=$Sc, Mg) is different from the case of Sr$_2$IrO$_4$ in which there is only one-band systems with the J=$\frac{1}{2}$ being half-filled, while the J=$\frac{3}{2}$ bands are completely filled \cite{AritaSr2IrO4,Zhang}. Then we believe Sr$_2$XOsO$_6$ compounds (X=Sc, Mg) might be considered as a multiband version of Sr$_2$IrO$_4$.

If the paramagnetic constrain is lifted we find a strong tendency to form magnetic moments increasing the electronic correlations. 
Then the present AFM insulating phase is then the consequence of such substantial correlation effects. 
We argue that the two and three correlated bands in the basis J,J$_z$ might be strongly related to the high magnetic ordering temperature of double perovskites Sr$_2$XOsO$_6$ (X$=$Sc, Mg) as it happens for other correlated half-filled materials in which the SOC is negligible \cite{Mravlje}. Further theoretical studies are needed to verify such hypothesis.

\section{Conclusions}

Using first-principles methods we investigate the electronic and magnetic structure of Sr$_2$XOsO$_6$ (X$=$Sc, Mg).
Our LDA+U+SOC calculations point out that the double peroskite Sr$_2$ScOsO$_6$ presents larger effects of electronic correlations than Sr$_2$MgOsO$_6$ while in this latter the effect of the SOC is dominant. The effect of the SOC combined with electronic correlations is to decrease the local magnetization M in both Sr$_2$XOsO$_6$ and to increase(descrease) the critical value of interaction strength U to get the insulating phase in Sr$_2$ScOsO$_6$(Sr$_2$MgOsO$_6$).
Within our LDA+U+SOC AFM calculations we identify the origin of the antiferromagnetic insulating phase to be related to the presence of two (X=Mg) and three (X=Sc) J=$\frac{1}{2},\frac{3}{2}$ half-filled bands splitted into fully filled lower and unfilled upper Hubbard bands.
We relate this AFM state to the presence of localized electrons close to the Mott state by performing LDA+DMFT and LDA+SOC+DMFT calculations.
The 5d$^2$ (X=Mg) and 5d$^3$ (X=Sc) electronic configurations in the LDA+DMFT produce respectively phases with promoted effects of electronic correlations far from and close to a Mott state. However when SOC is taken into account the effects of effective crystal-field splitting in the J,J$_z$ basis and electronic correlations shift Sr$_2$MgOsO$_6$ to be closer to a Mott state related to the J=$\frac{3}{2}$ bands while in Sr$_2$ScOsO$_6$ lead to a larger critical value of U for the Mott-Hubbard transition in J=$\frac{1}{2},\frac{3}{2}$ bands.
Our LDA+DMFT and LDA+SOC+DMFT calculations substantiate the importance of the inclusion of the SOC in investigating the electronic structure of such compounds and give evidence of correlated J=$\frac{1}{2},\frac{3}{2}$ bands in double perovskites Sr$_2$XOsO$_6$ (X$=$Sc, Mg), which might be candidates to host high magnetic ordering temperature. \\

\section{ACKNOWLEDGMENTS}
The author GG is indebted with M. Casula for the fruitfull discussions and thanks M. Aichhorn, A. Privitera for the careful and critical reading of the manuscript.


\begin{thebibliography}{99}

\bibitem{Retuerto} M. Retuerto, M.-R. Li, P. W. Stephens, J. Sánchez-Benítez, X. Deng, G. Kotliar, M. C. Croft, A. Ignatov, D. Walker, and M. Greenblatt, Chem. Mater., 2015, 27 (12), pp 4450–4458.

\bibitem{Philipp} J. B. Philipp, P. Majewski, L. Alff, A. Erb, R. Gross, T. Graf, M. S. Brandt, J. Simon, T. Walther, W. Mader, D. Topwal, and D. D. Sarma, Phys. Rev. B {\bf 68}, 144431 (2003).

\bibitem{Feng} H. L. Feng, M. Arai, Y. Matsushita, Y. Tsujimoto, Y. Guo, C. I. Sathish, X. Wang, Y.-H. Yuan, M. Tanaka, and K. Yamaura, J. Am. Chem. Soc., 2014, 136 (9), pp 3326–3329.

\bibitem{Krockenberger} Y. Krockenberger, K. Mogare, M. Reehuis, M. Tovar, M. Jansen, G. Vaitheeswaran, V. Kanchana, F. Bultmark, A. Delin, F. Wilhelm, A. Rogalev, A. Winkler, and L. Alff, Phys. Rev. B {\bf 75}, 020404(R) (2007).

\bibitem{Fukushima} T. Fukushima, A. Stroppa, S. Picozzi, J. M. Perez-Mato, Phys. Chem. Chem. Phys., 2011, 13, 12186-12190.

\bibitem{Kumar} S. Kumar, G. Giovannetti, J. van den Brink, S. Picozzi, Phys. Rev. B {\bf 82}, 134429 (2010).

\bibitem{Retuerto1} M. Retuerto, S. Skiadopoulou, M.-R. Li, A. M. Abakumov, M. Croft, A. Ignatov, T. Sarkar, B. M. Abbett, J. Pokorný, M. Savinov, D. Nuzhnyy, J. Prokleška, M. Abeykoon, P. W. Stephens, J. P. Hodges, P. Vaněk, C. J. Fennie, K. M. Rabe, S. Kamba, and M. Greenblatt, Inorg. Chem., 2016, 55 (9), pp 4320–4329.

\bibitem{Ou} X. Ou, Z. Li, F. Fan, H. Wang and H. Wu, Sc. Rep. 4, Article number: 7542 (2014).

\bibitem{Morrow1} R. Morrow, R. Mishra, O. D. Restrepo, M. R. Ball, W. Windl, S. Wurmehl, U. Stockert, B. B\"uchner, and P. M. Woodward , J. Am. Chem. Soc., 2013, 135 (50), pp 18824–18830.

\bibitem{Sr2ScOsO6} A. E. Taylor, R. Morrow, D. J. Singh, S. Calder, M. D. Lumsden, P. M. Woodward, and A. D. Christianson, Phys. Rev. B {\bf 91}, 100406(R) (2015).

\bibitem{Sr2MgOsO6} Y. Yuan, H. L. Feng, M. P. Ghimire, Y. Matsushita, Y. Tsujimoto, J. He, M. Tanaka, Y. Katsuya, and K. Yamaura, Inorg. Chem., 2015, 54 (7), pp 3422–3431 (2015).

\bibitem{ChenBalents} G. Chen and Leon Balents, Phys. Rev. B 84, 094420 (2011).

\bibitem{Pardo} V. Pardo and W.E. Pickett, Phys. Rev. B {\bf 80}, 054415 (2009).

\bibitem{Lee2} K.-W. Lee and W.E. Pickett, Phys. Rev. B {\bf 77}, 115101 (2008).

\bibitem{Gangopadhyay} S. Gangopadhyay and W.E. Pickett, Phys. Rev. B {\bf 93}, 155126 (2016).

\bibitem{GeorgesAnnuRev} A. Georges, L. de' Medici, and J. Mravlje, Annu. Rev. Condens. Matter Phys. {\bf 4} 137 (2013).

\bibitem{DeMedici} L. de' Medici, J. Mravlje, and A. Georges, Phys. Rev. Lett. {\bf 107}, 256401 (2011).

\bibitem{LDA} J. P. Perdew and A. Zunger, Phys. Rev. B {\bf 23}, 5048 (1981).

\bibitem{VASP} G. Kresse and J. Furthmuller, Phys. Rev. B {\bf 54}, 11 169 (1996); G. Kresse and J. Furthmuller, Comput. Mater. Sci. {\bf 6}, 15 (1996).

\bibitem{PAW} G. Kresse and D. Joubert, Phys. Rev. B {\bf 59}, 1758 (1999).

\bibitem{DFTplusU} A. I. Liechtenstein, V. I. Anisimov, and J. Zaanen. Phys. Rev. B {\bf 52}, R5467(R) (1995).

\bibitem{DMFT} A. Georges, G. Kotliar, W. Krauth, and M. J. Rozenberg, Rev. Mod. Phys. {\bf 68}, 13 (1996).

\bibitem{wannier90} A. A. Mostofi, J. R. Yates, Y. -S. Lee, I. Souza, D. Vanderbilt and N. Marzari, Comput. Phys. Commun. \textbf{178}, 685 (2008).

\bibitem{Caffarel} M. Caffarel and W. Krauth, Phys. Rev. Lett. {\bf 72}, 1545 (1994).

\bibitem{Capone} M. Capone, L. de' Medici, and A. Georges, Phys. Rev. B {\bf 76}, 245116 (2007).

\bibitem{Liebsch} A. Liebsch and H. Ishida, J. Phys.: Condens. Matter {\bf 24}, 053201 (2012).

\bibitem{2322} G. Giovannetti, L. de' Medici, M. Aichhorn, and M. Capone, Phys. Rev. B {\bf 91}, 085124 (2015).

\bibitem{BaCrO3GG} Gianluca Giovannetti, Markus Aichhorn, and Massimo Capone, Phys. Rev. B {\bf 90}, 245134  (2014)

\bibitem{ARPACK} R. B. Lehoucq, D. C. Sorensen, and C. Yang, ARPACK User's Guide (SIAM, Philadelphia, 1997).

\bibitem{AritaSr2IrO4} R. Arita, J. Kunes, A. V. Kozhevnikov, A. G. Eguiluz, and M. Imada Phys. Rev. Lett. {\bf 108}, 086403 (2012).

\bibitem{notaBZ} The high symmetry direction of the Brillouin zone are defined as $\Gamma$=(0,0,0), X=(0.5,0,0), Y=(0,0.5,0), M=(0.5,0.5,0) and Z=(0,0,0.5).

\bibitem{Lee} K.-W. Lee and W. E. Pickett, EPL, 80, 3 (2007).

\bibitem{notaAFM} In our LDA+U+SOC calculations in the type-I ordered magnetic phase no large differences in spectral distributions are found in how the spins are oriented in the ab-plane.

\bibitem{Sasioglu} {Ersoy \ifmmode \mbox{\c{S}}\else \c{S}\fi{}a\ifmmode \mbox{\c{s}}\else \c{s}\fi{}\ifmmode \imath \else \i \fi{}o\ifmmode \breve{g}\else \u{g}\fi{}lu,  Christoph Friedrich andStefan Bl\"ugel}, Phys. Rev. B 83, 121101(R) (2011).

\bibitem{LiOsO3GGMC} G. Giovannetti and M. Capone, Phys. Rev. B {\bf 90}, 195113 (2014).

\bibitem{LiOsO3LiNbO3} D. Puggioni, G. Giovannetti, M. Capone, and J. M. Rondinelli, Phys. Rev. Lett. {\bf 115}, 087202 (2015)

\bibitem{Jung} M.-C. Jung, Young-Joon Song, Kwan-Woo Lee, and Warren E. Pickett, Phys. Rev. B {\bf 87}, 115119 (2013).

\bibitem{notaPade} The orbital-resolved spectral density of states obtained within DMFT scheme is obtained using the Pad\'e approximation starting from the imaginary-frequency Self-energies.

\bibitem{Zhang} H. Zhang, K. Haule, and D. Vanderbilt, Phys. Rev. Lett. {\bf 111}, 246402 (2013).

\bibitem{Mravlje} J. Mravlje, M. Aichhorn, and A. Georges, Phys. Rev. Lett. {\bf 108}, 197202 (2012).

\end{thebibliography}
\end{document}